\documentclass[12pt,a4paper]{article}

\usepackage{xcolor}
\newcommand{\rev}[1]{\textcolor{black}{#1}}

\usepackage{array}

\usepackage[T1]{fontenc}
\usepackage[utf8]{inputenc}
\usepackage{lmodern}
\usepackage[top=2.5cm, bottom=2.5cm, left=2.8cm, right=2.8cm]{geometry}
\usepackage{amsmath}
\usepackage[version=4]{mhchem}
\usepackage{graphicx}
\usepackage{booktabs}
\usepackage{multirow}
\usepackage{caption}
\usepackage{ragged2e}
\usepackage[numbers,sort&compress]{natbib}
\usepackage[colorlinks=true, linkcolor=blue!50!black,
            citecolor=blue!50!black, urlcolor=blue!50!black]{hyperref}
\usepackage{orcidlink}

\usepackage{titlesec}
\titleformat{\section}{\large\bfseries}{\thesection.}{0.5em}{}
\titleformat{\subsection}{\normalsize\bfseries}{\thesubsection.}{0.5em}{}

\usepackage{float}
\begin{document}

\begin{justify}

{\LARGE\bfseries
van der Waals Nanogap Loading of a
\ce{TiO2} Nanobar Metasurface for Dual-Channel Fano Refractive Index Sensing
\par}
\end{justify}


\begin{flushleft}
{\large Shoumik Debnath\,\orcidlink{0009-0001-9500-5319} and Sudipta Saha\,\orcidlink{0009-0003-9110-8059}$^{*}$\par}

\vspace{0.5em}

{\small
Department of Electrical and Electronic Engineering,\\
Bangladesh University of Engineering and Technology,\\
Dhaka 1205, Bangladesh\\[3pt]
$^{*}$E-mail:
\href{mailto:sudiptasaha@ari.buet.ac.bd}{sudiptasaha@ari.buet.ac.bd}
\par}
\end{flushleft}

\noindent\textbf{Abstract.}\;
Filling the gap of a \ce{TiO2} nanobar-pair metasurface with
$\alpha$-\ce{MoO3}, a biaxial orthorhombic crystal, produces two
high-$Q$ Fano resonances with \rev{unequal} quality factors:
$Q_\mathrm{TE} = 87$ at 863.3\,nm and $Q_\mathrm{TM} = 31$ at
960.1\,nm, separated by 97\,nm.
\rev{The same device with amorphous \ce{Sb2S3}, which is isotropic,
gives $Q_\mathrm{TE}=66$ and $Q_\mathrm{TM}=20$, a ratio of 3.35
against 2.85 for $\alpha$-\ce{MoO3}, so the inequality of the two
quality factors is not specific to the biaxial crystal and
originates instead in the in-plane anisotropy of the resonator
geometry.
An unequal pair of quality factors is therefore not by itself
evidence of material anisotropy.
What the biaxial fill controls at fixed geometry is the partitioning
of the response between the two channels: each polarization samples a
different principal index, so the two resonance positions, the two
linewidths and the ratio of the two sensitivities respond to a single
material choice in different proportions.}
The TE channel delivers sensitivity $S=155.3$\,nm\,RIU$^{-1}$,
figure of merit $15.71$\,RIU$^{-1}$, and \rev{an ideal refractive
index resolution of} $6.44\times10^{-5}$\,RIU; TM delivers
$S=139.1$\,nm\,RIU$^{-1}$, $4.44$\,RIU$^{-1}$, and
$7.19\times10^{-5}$\,RIU.
Simultaneous readout produces a \rev{fixed polarization response} with
isotropic slope 0.896\rev{, which the gap material changes: the same
geometry filled with amorphous \ce{Sb2S3} gives 0.620}.
\rev{All results reported here are numerical; no device was
fabricated and no measurement was performed.}

\section{Introduction}

Refractive index (RI) sensing detects biomolecular interactions
without labels or amplification across clinical diagnostics, drug
discovery, food safety, and environmental
monitoring.\cite{Homola2008SPR,Altug2023NanoReview}
Surface plasmon resonance (SPR) has served as the laboratory
standard, but ohmic losses in the metal film cap quality factors
below $Q\sim 100$, thermal drift introduces baseline instability,
and single-channel readout cannot decouple specific binding from
bulk RI drift without a separate reference
arm.\cite{Homola2008SPR}
All-dielectric metasurfaces eliminate the loss constraint:
electromagnetic resonances arise from Mie-type electric and magnetic
dipole modes,\cite{Evlyukhin2012,Staude2013} \rev{which avoid the
ohmic absorption of metals and so give} access to quality factors
far beyond what metals
allow.\cite{Kuznetsov2016Science,Koshelev2021ACSPhoton,Chen2016review}

Among dielectric resonances, quasi-bound states in the continuum
(q-BICs) are a widely used high-$Q$ mechanism in metasurface
sensors.
A symmetry perturbation in the unit cell couples an otherwise dark
BIC mode to the far field,\cite{Hsu2016BIC,Koshelev2018PRL,%
Campione2016ACS} producing a Fano lineshape in transmission whose
linewidth narrows quadratically as the perturbation decreases.
$Q$ is therefore a freely tunable parameter at fixed
geometry,\cite{Lawrence2020disorder} and q-BIC sensors have reached
sensitivities above 200\,nm\,RIU$^{-1}$ and figures of merit
exceeding 50\,RIU$^{-1}$.\cite{Yesilkoy2019dielectric,Tittl2018Science,%
Jing2023quasiBIC,Li2025quasiBIC,Zhao2026QBIC}
\rev{The structure studied here does not employ that mechanism: its
two resonators are identical and the gap fill is centred, so the unit
cell retains its mirror symmetries and no symmetry-protected dark
mode is rendered radiative by the fill (Section~3.1).}

Almost all reported sensors read a single resonance per measurement.
A second channel is needed for two reasons specific to real
biosensing.
First, when two molecular species bind simultaneously, a single
resonance shift cannot separate them.
Second, biologically ordered films---lipid bilayers, collagen
fibres, DNA monolayers---present different RIs along and across
their ordering axis; a single-polarization sensor discards this
anisotropy as noise, whereas simultaneous TE and TM readout
encodes it in the ratio of the two
shifts.\cite{Kramadhati2025SOP,Butt2026polarization}
Dual-channel metasurfaces have been reported using distinct
resonators per channel,\cite{Yang2025Si3N4,Wang2024dualFano,%
Vijaya2023dual} but multi-element unit cells introduce alignment
tolerances and typically restrict both channels to comparable $Q$
values because geometry, not material, drives both.

A material-controlled route \rev{offers a different set of degrees of
freedom}.
An isotropic fill presents the same \rev{permittivity} to both
polarizations\rev{, though the two channels may still respond to it
unequally because the resonator geometry is itself anisotropic in
plane}; a biaxial crystal presents a different
\rev{principal index} to each, so \rev{the same material
substitution acts on the two channels unequally and the balance
between them becomes a material parameter at fixed geometry}.
Here we \rev{examine this numerically} using $\alpha$-\ce{MoO3}
(biaxial, orthorhombic) and amorphous and crystalline \ce{Sb2S3}
(both isotropic) as \rev{reference fills, and we report both what the
biaxial fill does and does not account for}.

\section{Device Design and Simulation}

\subsection{Geometry}

\begin{figure}[tbp]
\centering
\includegraphics[width=0.88\textwidth]{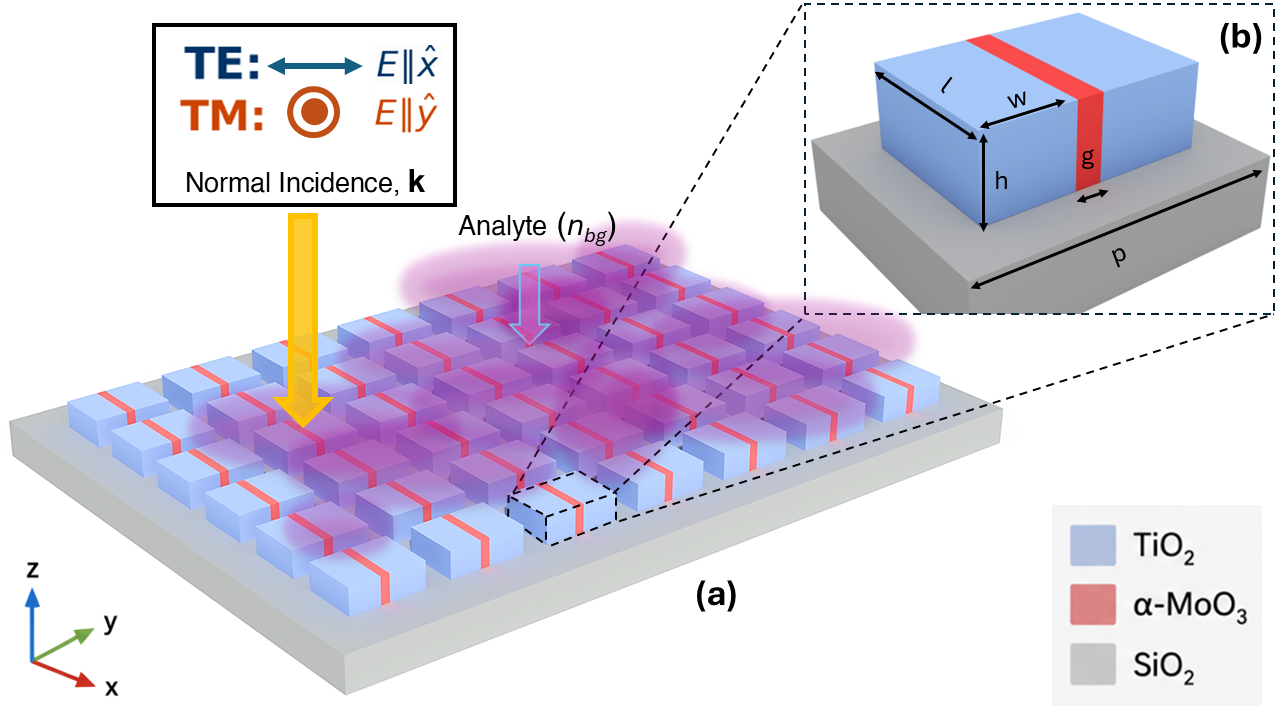}
\caption{Device schematic.
\textbf{(a)} Perspective view of the \ce{TiO2} metasurface.
The 60\,nm gap is filled with $\alpha$-\ce{MoO3},
amorphous \ce{Sb2S3}, or crystalline \ce{Sb2S3}.
TE ($E\!\parallel\!\hat{x}$) and TM ($E\!\parallel\!\hat{y}$)
channels are accessed by rotating the incident polarization;
the analyte medium $n_\mathrm{bg}$ surrounds the structure.
\textbf{(b)} Unit cell: $W=120$\,nm, $L=450$\,nm, $H=250$\,nm,
$g=60$\,nm, $p=600$\,nm.}
\label{fig:schematic}
\end{figure}

The unit cell (Fig.\,\ref{fig:schematic}) consists of two \ce{TiO2}
nanobars ($W\times L\times H = 120\times 450\times 250$\,nm) aligned
along $\hat{y}$, separated by a 60\,nm gap, and arranged on a
\ce{SiO2} substrate ($n_s=1.46$) with period $P=600$\,nm.
The geometry is fixed across all three fill experiments.
\ce{TiO2} has refractive index $n=2.35$ at 850--970\,nm, consistent
with plasma-assisted ALD
films,\cite{Siefke2016TiO2,Kischkat2012,Garcia2021TiO2,Profijt2011ALD}
and is established in all-dielectric RI
sensors.\cite{Abbas2020TiO2,Wu2022TiO2GMR}
\rev{The unit cell is itself strongly anisotropic in plane: the
resonators are 120\,nm wide and 450\,nm long, separated by 60\,nm
along $x$ and by 150\,nm along $y$ within the 600\,nm period.
This is relevant to the interpretation of the results in
Section~3.1.}

\subsection{Gap fill materials}

\begin{figure}[tbp]
\centering
\includegraphics[width=\textwidth]{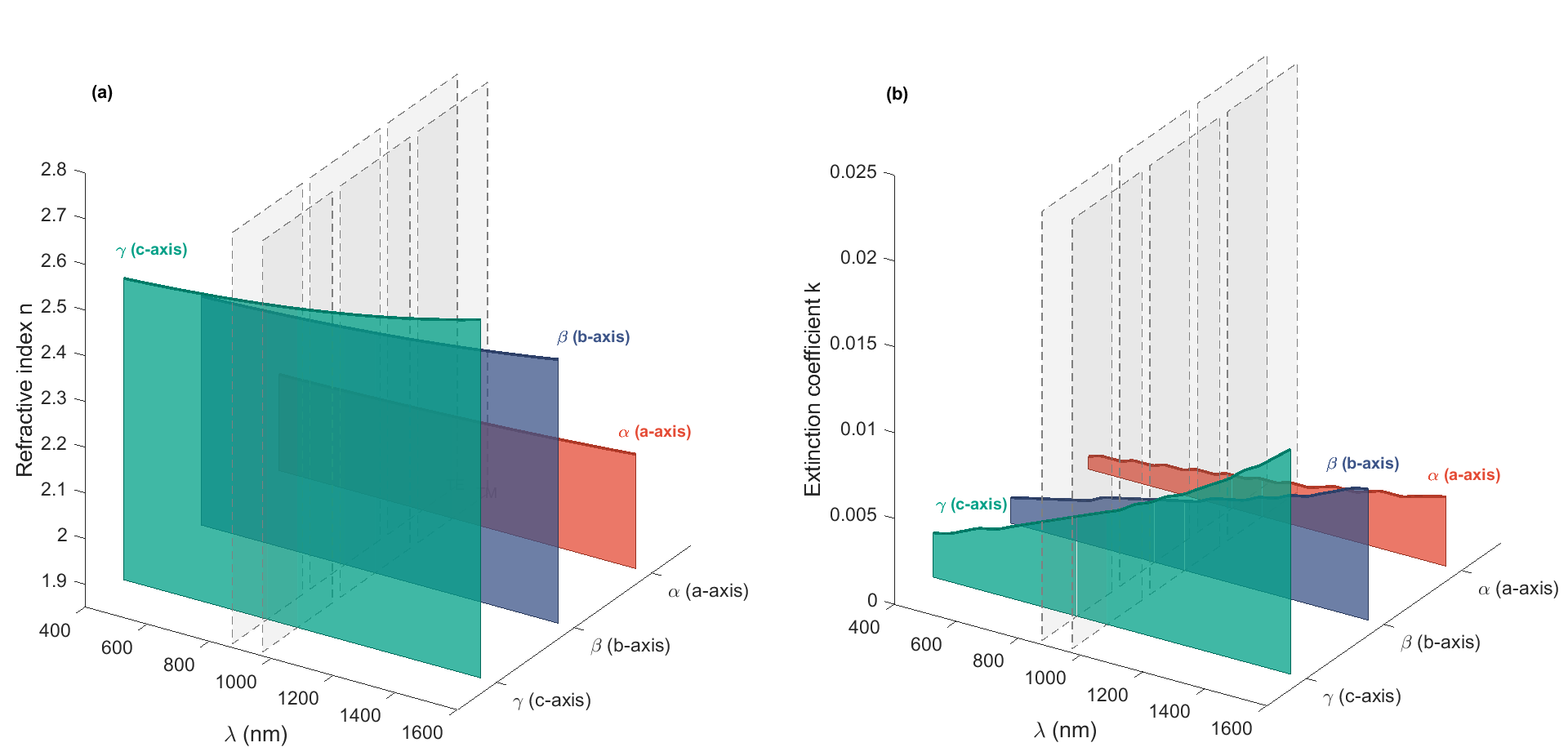}
\caption{Optical constants of $\alpha$-\ce{MoO3} (Lajaunie
\textit{et al.}\cite{Lajaunie2013}).
\textbf{(a)} Refractive index $n(\lambda)$: three distinct values
from inequivalent Mo--O bond environments.
\textbf{(b)} Extinction coefficient $k(\lambda)$: all axes satisfy
$k<0.015$ in the sensor window (860--980\,nm, shaded).}
\label{fig:nk}
\end{figure}

$\alpha$-\ce{MoO3} (orthorhombic, \textit{Pbnm}) has three
inequivalent Mo--O bond environments along its $a$, $b$, and $c$
axes, giving principal indices $n_\alpha=2.092$, $n_\beta=2.411$,
and $n_\gamma=2.607$ at the operating
wavelengths.\cite{Lajaunie2013}
Fig.\,\ref{fig:nk} shows the full spectral dependence.
All three extinction coefficients satisfy $k<0.015$ at 860--980\,nm\rev{.
This bound alone does not establish whether the resonances are
radiatively or absorptively limited; separating the two contributions
requires a direct comparison of lossless and lossy simulations, which
is not evaluated by the simulations reported here}.
The material was assigned in \rev{the simulations} as a diagonal
anisotropic medium with $\varepsilon_{xx}=n_\beta^2$,
$\varepsilon_{yy}=n_\alpha^2$, and $\varepsilon_{zz}=n_\gamma^2$,
where $x$ crosses the gap, $y$ aligns with the bars, and $z$ is the
bar height.
\rev{This assignment is a modelling choice adopted to place the two
largest in-plane index contrasts in the plane of the device, and we
state it as such.
In conventionally exfoliated $\alpha$-\ce{MoO3} the van der Waals
stacking direction lies normal to the flake, so the two in-plane
directions would instead be the remaining pair of crystallographic
axes.
Under that alternative assignment the axis-pair contrast sampled by
TE is unchanged at 0.983 and that sampled by TM becomes 1.437 in
place of 2.420, so the two channels remain inequivalent.
This follows from the permittivity tensor alone; the alternative
assignment was not re-simulated, and the resonance positions,
linewidths and sensitivities reported below apply to the assignment
used here.}

\begin{figure}[tbp]
\centering
\includegraphics[width=\textwidth]{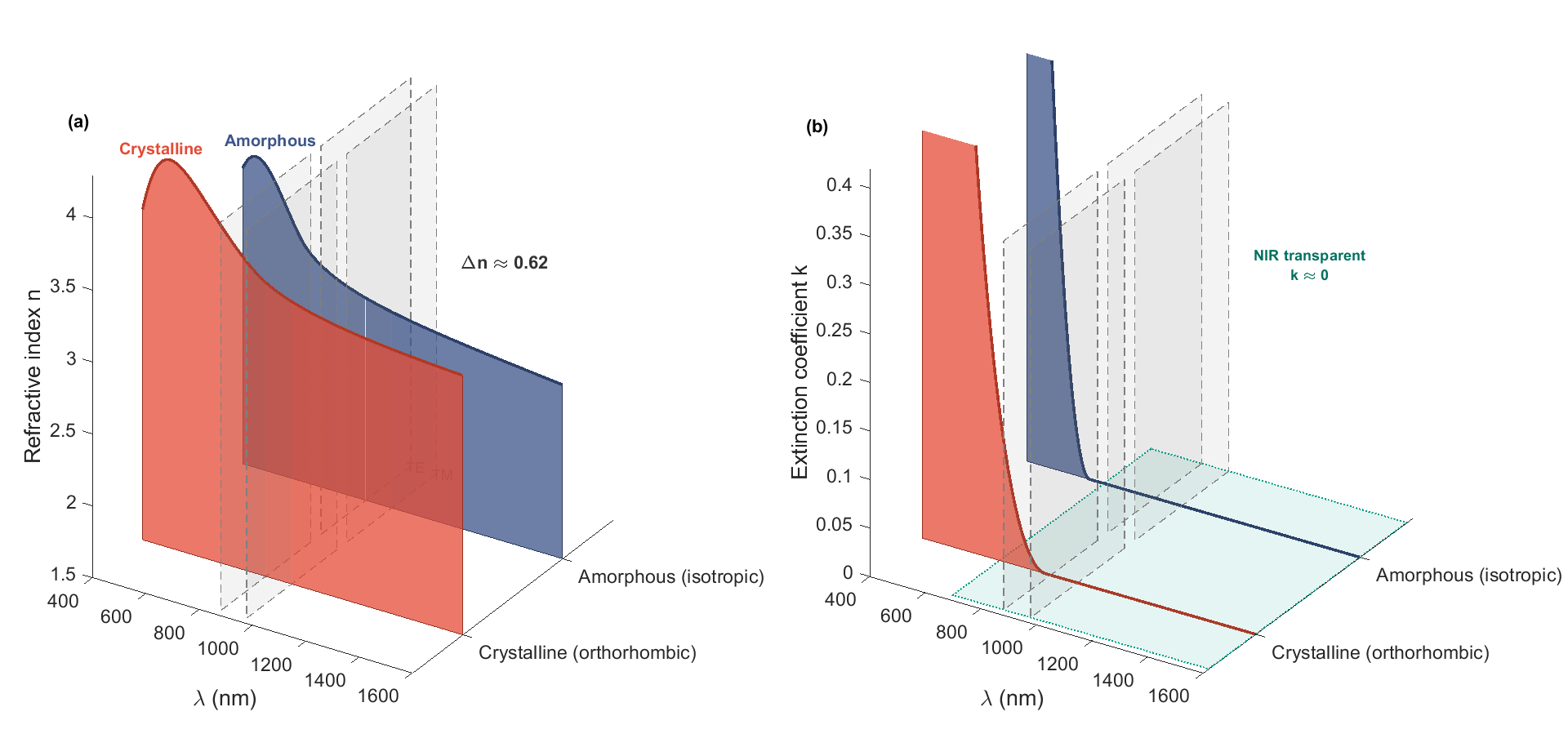}
\caption{Optical constants of \ce{Sb2S3}.\cite{Delaney2020}
\textbf{(a)} Refractive index: amorphous ($n=2.90$) and crystalline
($n=3.52$) phases.
\textbf{(b)} Extinction coefficient: both phases satisfy
$k\approx 0$ above 860\,nm (shaded).}
\label{fig:sb2s3_nk}
\end{figure}

Amorphous and crystalline \ce{Sb2S3} served as isotropic
reference fills with $n=2.90$ and $n=3.52$ respectively, both
with $k\approx 0$ above 860\,nm
(Fig.\,\ref{fig:sb2s3_nk}).\cite{Delaney2020}
The two phases supply isotropic indices of 2.90 and 3.52 that bracket
the $\alpha$-\ce{MoO3} principal values, giving a broad-based isotropic
reference across the full index range sampled by the biaxial fill.

\subsection{FDTD setup}

All simulations used Lumerical FDTD\cite{LumericalFDTD} following
standard FDTD methodology.\cite{Taflove2005FDTD}
\rev{The computational domain is a single $600\times600$\,nm unit
cell.}
Periodic boundary conditions were applied in $x$ and $y$; PML
absorbers\cite{Berenger1994PML} bounded the domain in $z$.
A broadband plane wave (700--1100\,nm) was launched at normal
incidence at 0\textdegree{} (TE) or 90\textdegree{} (TM), with the
analyte medium filling the full superstrate including bar surfaces
and gap interior.
\rev{The source is injected from the analyte side and propagates towards
the substrate; a polarization angle of 0\textdegree{} gives
$E\!\parallel\!\hat{x}$ across the gap and 90\textdegree{} gives
$E\!\parallel\!\hat{y}$ along the bars.
Transmission was recorded by a frequency-domain power monitor placed
beyond the substrate and normalised to the source, with a second
monitor recording reflection above the source plane, and the spectral
window was sampled at 3000 points, a spacing of 0.133\,nm.}
Six values $n_\mathrm{bg}=1.00$--$1.05$ (step 0.01) were simulated
per polarization per fill (36 runs total).
\rev{The mesh was non-uniform, with a 2\,nm step through the resonator
and gap region coarsening away from the structure.
We note that a systematic mesh convergence study was not carried out
for the results reported here.}

\rev{All materials were modelled as non-dispersive over the operating
window using the fixed refractive indices given above, namely
$n_\alpha=2.092$, $n_\beta=2.411$ and $n_\gamma=2.607$ for
$\alpha$-\ce{MoO3}, $n=2.90$ and $n=3.52$ for the two \ce{Sb2S3}
phases, $n=2.35$ for \ce{TiO2} and $n=1.46$ for the \ce{SiO2}
substrate; Figs.\,\ref{fig:nk} and \ref{fig:sb2s3_nk} show the
dispersion from which those values are taken and justify that
treatment over the wavelength range used.
The gap fill occupies the 60\,nm gap over the full bar height and is
flush with the bar tops.}

\rev{Resonance parameters were extracted by fitting the transmission
to the Fano lineshape
\begin{equation}
T(\lambda) = T_\mathrm{bg}\,
\frac{(q+\epsilon)^2}{1+\epsilon^2},
\qquad
\epsilon = \frac{2(\lambda-\lambda_0)}{\gamma},
\label{eq:fano}
\end{equation}
by bounded nonlinear least squares in the four parameters
$\lambda_0$, $\gamma$, $q$ and $T_\mathrm{bg}$, with the quality
factor obtained as $Q=\lambda_0/\gamma$ and the coefficient of
determination $R^2$ reported for every fit.
The fitting window is 855--900\,nm for TE and 925--985\,nm for TM of
the $\alpha$-\ce{MoO3} device, and 878--915\,nm and 955--1035\,nm
respectively for amorphous \ce{Sb2S3}.
We state the window for every reported fit because, as described in
Section~3.1, an inappropriately wide window was the source of an
error in the earlier analysis of the \ce{Sb2S3} data.}

\section{Results and Discussion}

\subsection{\rev{Origin of the channel inequality}}

\rev{Two distinct permittivity contrasts govern different aspects of
this device, and they are kept separate here.
The first is the contrast between the gap fill and the resonator
material,
\begin{equation}
\Delta\varepsilon_\mathrm{fb}
= \bigl|\,n_\mathrm{fill}^2 - n_\mathrm{bar}^2\,\bigr|,
\label{eq:dfb}
\end{equation}
evaluated for the principal index sampled by the incident
polarization, which is non-zero for every fill considered here.
The second is the in-plane contrast between the two principal axes of
the fill,
\begin{equation}
\delta_\mathrm{a} =
\bigl|\,\varepsilon_{xx}-\varepsilon_{yy}\,\bigr|,
\label{eq:daxis}
\end{equation}
which is identically zero for an isotropic fill and equal to 1.437
for $\alpha$-\ce{MoO3} in the assignment of Section~2.2.
Table~\ref{tbl:contrast} lists both for the three fills.
Conflating the two led to the inconsistency that an isotropic fill
would give $\Delta\varepsilon=0$ and hence an unbounded quality
factor, which is incompatible with the finite quality factors
obtained for the \ce{Sb2S3} devices.}

\begin{table}[tbp]
\centering
\caption{\rev{Permittivity contrasts for the three gap fills.
$\Delta\varepsilon_\mathrm{fb}$ is evaluated against
$n_\mathrm{bar}=2.35$ for the principal index sampled by each
polarization; $\delta_\mathrm{a}$ is the in-plane axis-pair
contrast of the fill.}}
\label{tbl:contrast}
\small
\begin{tabular}{lccc}
\toprule
\rev{Gap fill} & \rev{$\Delta\varepsilon_\mathrm{fb}$ (TE)}
  & \rev{$\Delta\varepsilon_\mathrm{fb}$ (TM)}
  & \rev{$\delta_\mathrm{a}$} \\
\midrule
\rev{$\alpha$-\ce{MoO3} (biaxial)}    & \rev{0.290} & \rev{1.146} & \rev{1.437} \\
\rev{a-\ce{Sb2S3} (isotropic)}        & \rev{2.888} & \rev{2.888} & \rev{0} \\
\rev{c-\ce{Sb2S3} (isotropic)}        & \rev{6.868} & \rev{6.868} & \rev{0} \\
\bottomrule
\end{tabular}
\end{table}

\begin{figure}[tbp]
\centering
\includegraphics[width=\textwidth]{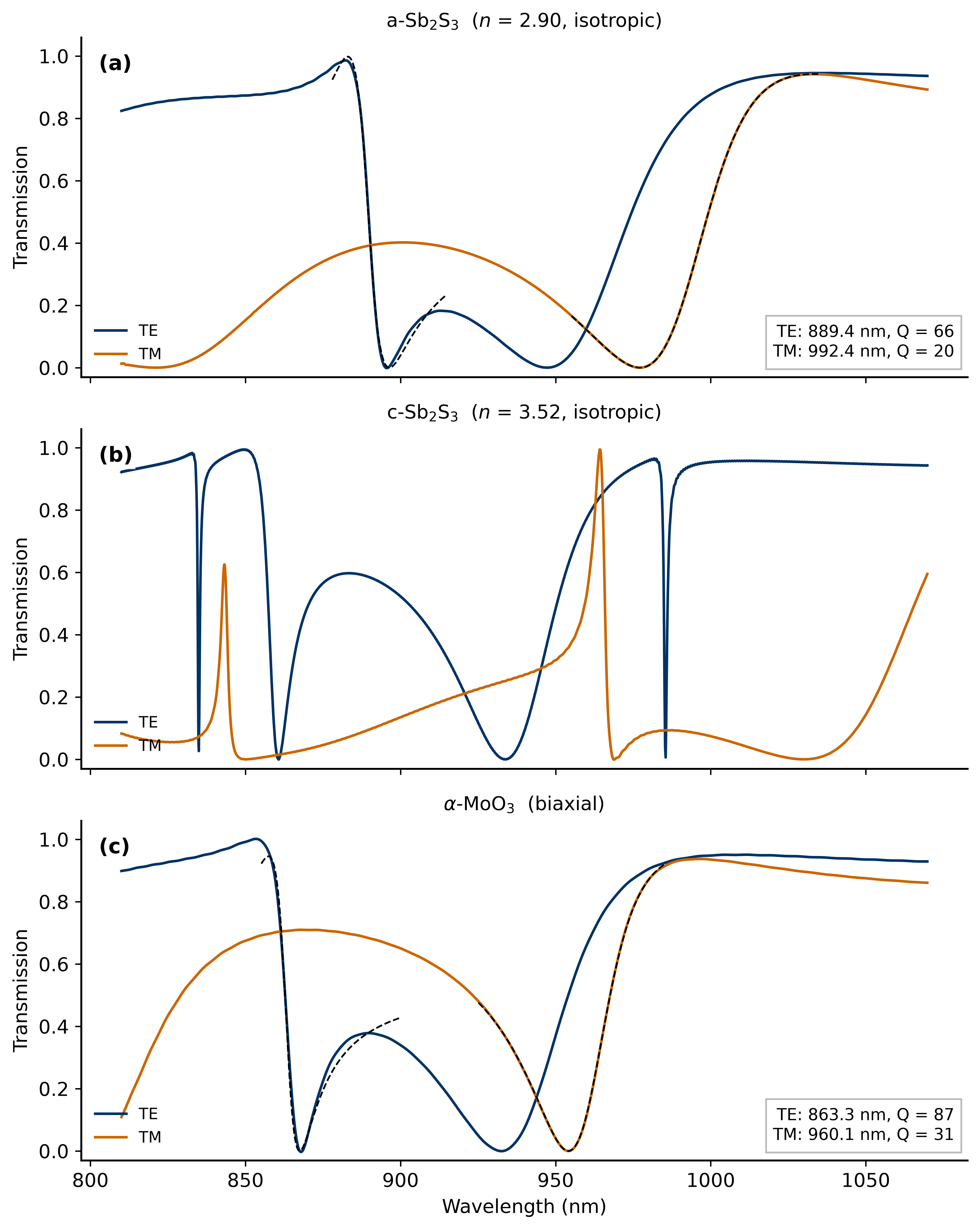}
\caption{Fano resonance \rev{spectra} across three gap fills.
\textbf{(a,\,b)} Isotropic \ce{Sb2S3} fills.
\textbf{(c)} $\alpha$-\ce{MoO3} biaxial fill: \rev{resonances} at
863\,nm ($Q=87$, TE) and 960\,nm ($Q=31$, TM), separated by 97\,nm.
\rev{Unequal TE and TM linewidths are present for all three fills;
fitted parameters are given in Table~\ref{tbl:controls}.}}
\label{fig:activation}
\end{figure}

Fig.\,\ref{fig:activation} shows transmission for all three fills.
With $\alpha$-\ce{MoO3} \rev{(panel c)}, TE yields a sharp Fano
resonance at 863\,nm ($Q=87$) and TM a broader one at 960\,nm
($Q=31$), separated by 97\,nm.
\rev{The two \ce{Sb2S3} devices (panels a, b) also give two
spectrally separated resonances with unequal linewidths;
Table~\ref{tbl:controls} reports the fitted parameters.}

\begin{table}[tbp]
\centering
\caption{\rev{Fitted resonance parameters at $n_\mathrm{bg}=1.00$
for the biaxial fill and the amorphous isotropic reference, each
resonance fitted within its own spectral window.}}
\label{tbl:controls}
\small
\begin{tabular}{llcccc}
\toprule
\rev{Gap fill} & \rev{Channel} & \rev{$\lambda_0$ (nm)} & \rev{$Q$}
  & \rev{FWHM (nm)} & \rev{$R^2$} \\
\midrule
\rev{$\alpha$-\ce{MoO3}} & \rev{TE} & \rev{863.32} & \rev{87.3} & \rev{9.89}  & \rev{0.983} \\
\rev{$\alpha$-\ce{MoO3}} & \rev{TM} & \rev{960.14} & \rev{30.7} & \rev{31.32} & \rev{1.000} \\
\rev{a-\ce{Sb2S3}}       & \rev{TE} & \rev{889.36} & \rev{66.4} & \rev{13.40} & \rev{0.997} \\
\rev{a-\ce{Sb2S3}}       & \rev{TM} & \rev{992.40} & \rev{19.8} & \rev{50.09} & \rev{1.000} \\
\bottomrule
\end{tabular}
\end{table}

\rev{The isotropic reference therefore also exhibits an asymmetric
quality-factor ratio, $Q_\mathrm{TE}/Q_\mathrm{TM}=3.35$, which is
larger than the value of 2.85 obtained with the biaxial fill and has
the same ordering of the two channels.
The inequality $Q_\mathrm{TE}\neq Q_\mathrm{TM}$ is therefore not
specific to the biaxial crystal.
Fitting these spectra over a single wide window, 830--1050\,nm for TE,
merges the two resonances of the amorphous device near 895 and
947\,nm and returns an artificially broad linewidth of 99.5\,nm with
$R^2=0.85$; each resonance is fitted separately here for that reason.}

\rev{The control therefore carries a methodological conclusion in place
of the one previously drawn from it.
An unequal pair of TE and TM quality factors is not by itself evidence
of material anisotropy, since here the isotropic fill produces the
larger asymmetry of the two.
Separating the polarization asymmetry that the geometry produces from
the tuning that the fill contributes requires an isotropic reference of
this kind.
The asymmetry observed here is already present in the isotropic
control, and therefore does not require material anisotropy; we do not
attempt to apportion it quantitatively between the two contributions.}

\rev{The origin of the channel inequality is the in-plane anisotropy
of the unit cell itself (Section~2.1).
The TE and TM channels couple to different modes of a geometry that
is 120\,nm wide and 450\,nm long, with different field distributions
and different resonance linewidths, for any gap fill.
Consistent with this, the device retains both of its mirror planes
when the principal axes of the fill are aligned with the structural
axes, so no symmetry-protected dark mode is rendered radiative by the
fill and the resonances are not symmetry-broken quasi-BICs.
We describe them instead as Fano-shaped leaky resonances of the
periodic structure.}

\rev{What the biaxial fill does control at fixed geometry is the
partitioning of the response between the two channels.
Because each polarization samples a different principal index, a
single material substitution displaces the two resonance positions
and the two linewidths in different proportions, and it sets the
ratio of the two channel sensitivities: 0.896 for
$\alpha$-\ce{MoO3} against 0.620 for amorphous \ce{Sb2S3}, from an
unchanged geometry (Sections~3.3 and 3.4).
A quantitative relationship between $\delta_\mathrm{a}$ and the two
quality factors is not established by the three fills reported here,
and we do not propose one; determining it requires a continuous sweep
of the in-plane anisotropy at fixed $\Delta\varepsilon_\mathrm{fb}$.}

\begin{figure}[h]
\centering
\includegraphics[width=\textwidth]{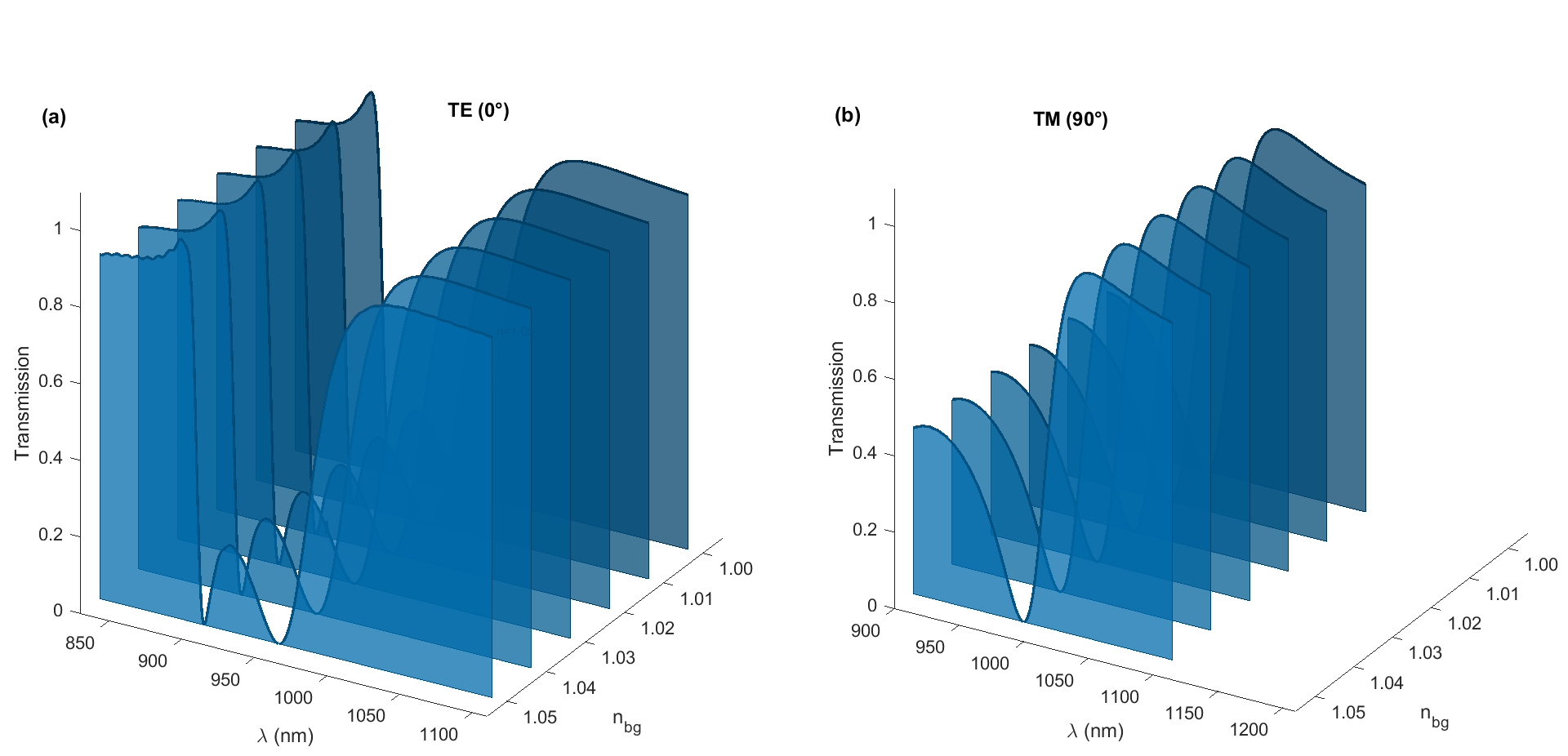}

\includegraphics[width=\textwidth]{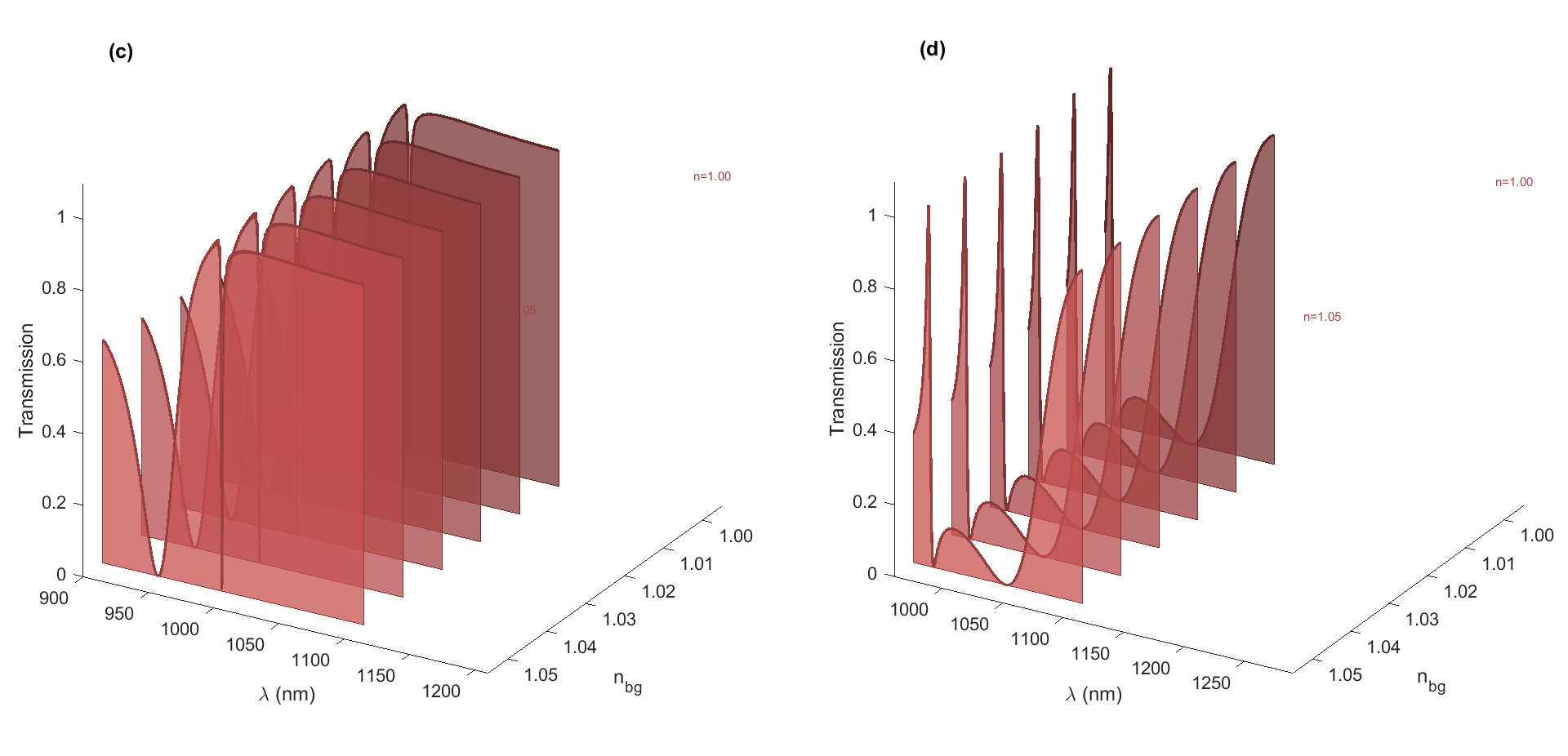}

\caption{\ce{Sb2S3} transmission spectra, $n_\mathrm{bg}=1.00$--$1.05$.
Each panel shows TE (left) and TM (right).
\textbf{(a), (b)} Amorphous \ce{Sb2S3} ($n=2.90$)\rev{.}
\textbf{(c), (d)} Crystalline \ce{Sb2S3} ($n=3.52$): higher index
activates multiple overlapping TM resonances absent in TE.}
\label{fig:sb2s3_waterfalls}
\end{figure}

The \ce{Sb2S3} RI sweeps (Fig.\,\ref{fig:sb2s3_waterfalls}) are consistent with
the pattern across $n_\mathrm{bg}=1.00$--$1.05$.
Amorphous \ce{Sb2S3} (panel a) produces monotonically shifting
resonances \rev{in both channels, with $S=153.1$\,nm\,RIU$^{-1}$ for
TE and $94.9$\,nm\,RIU$^{-1}$ for TM}.
Crystalline \ce{Sb2S3} (panel b) shifts resonances to longer
wavelengths and activates multiple overlapping guided-mode features
in TM that are absent in TE, producing a multi-peak spectrum
unsuitable for single-peak RI tracking.
\rev{Several of these features are individually sharp: fitted in
isolation they give $Q$ of approximately $1.1\times10^{3}$ at
834.9\,nm and $8.6\times10^{2}$ at 985.3\,nm in TE, and
$2.7\times10^{2}$ at 965.2\,nm in TM.
The first two have linewidths near the 0.133\,nm spectral sampling
used here and should be read as lower bounds on sharpness rather than
converged values.
The limitation of the crystalline phase is therefore not an absence
of high-$Q$ features but the number of them within the operating
window.}

\subsection{Fano Resonance Characterisation of the
\texorpdfstring{$\alpha$-\ce{MoO3}}{alpha-MoO3} Device}

\begin{figure}[tbp]
\centering
\includegraphics[width=\textwidth]{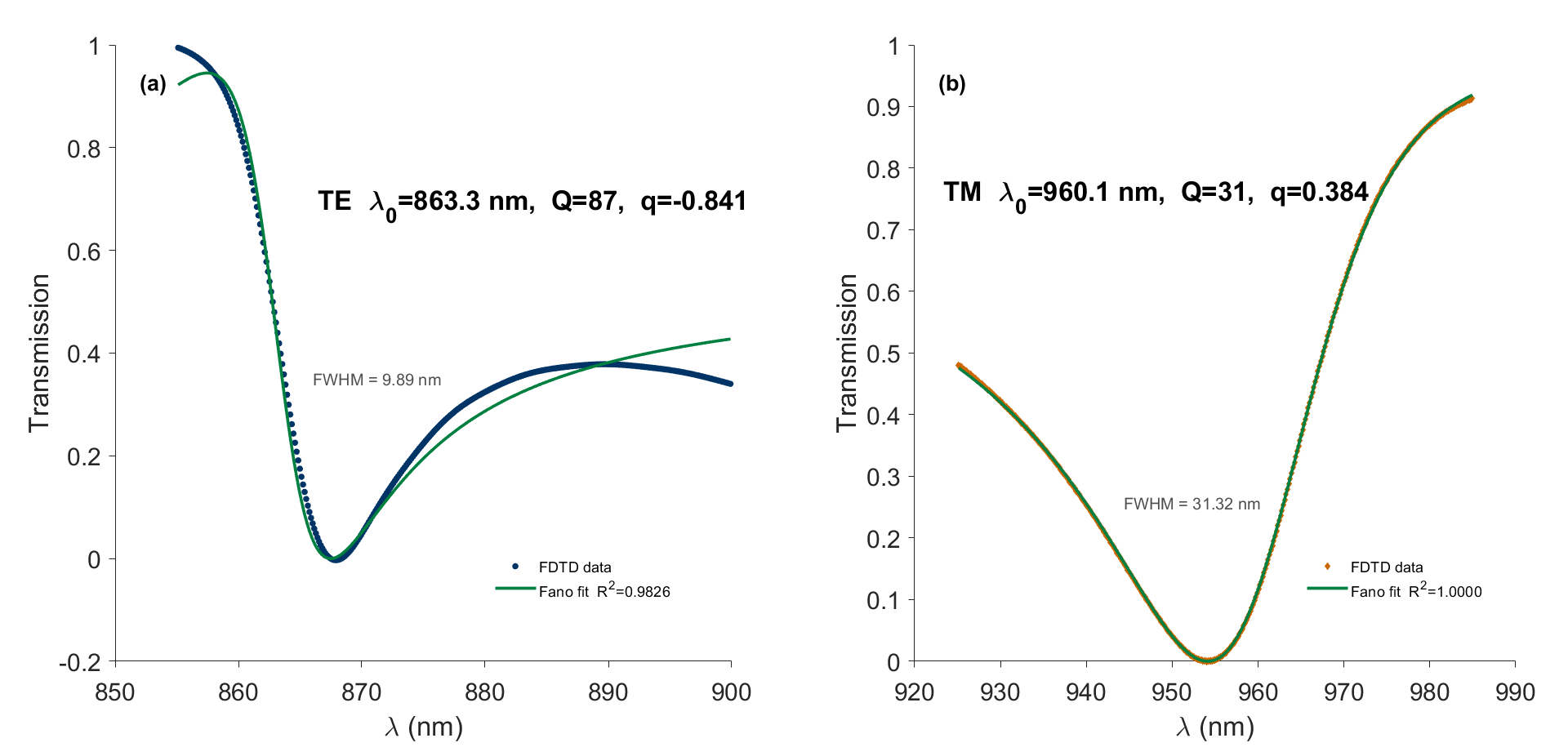}
\caption{Fano resonance fits at $n_\mathrm{bg}=1.00$,
$\alpha$-\ce{MoO3} fill.
\textbf{(a)} TE: $\lambda_0=863.3$\,nm, $Q=87$, $q=-0.841$,
FWHM\,$=9.89$\,nm, $R^2=0.983$.
\textbf{(b)} TM: $\lambda_0=960.1$\,nm, $Q=31$, $q=+0.384$,
FWHM\,$=31.32$\,nm, $R^2=1.000$.
Filled symbols: FDTD data. Solid curves: Fano fits.}
\label{fig:fano}
\end{figure}

\begin{figure}[t]
\centering
\begin{minipage}[t]{0.48\textwidth}
  \centering
  \includegraphics[width=\textwidth]{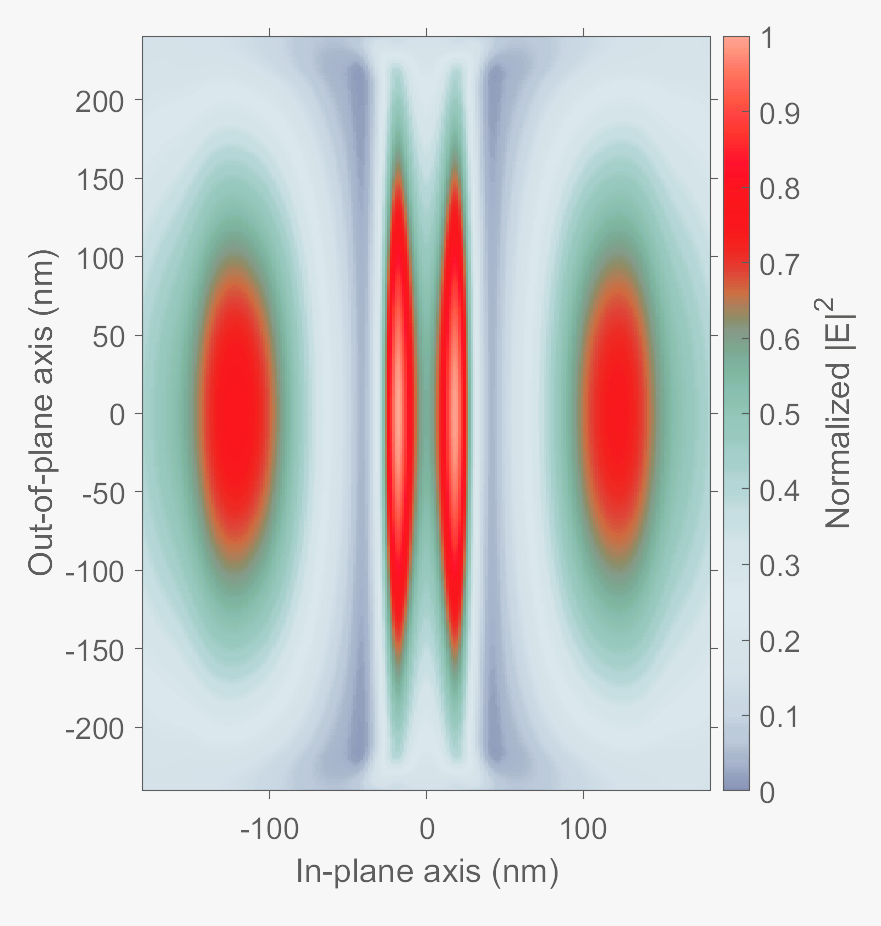}
  \vspace{2pt}\\
  \small\textbf{(a) TE, 0\textdegree}
\end{minipage}%
\hfill
\begin{minipage}[t]{0.48\textwidth}
  \centering
  \includegraphics[width=\textwidth]{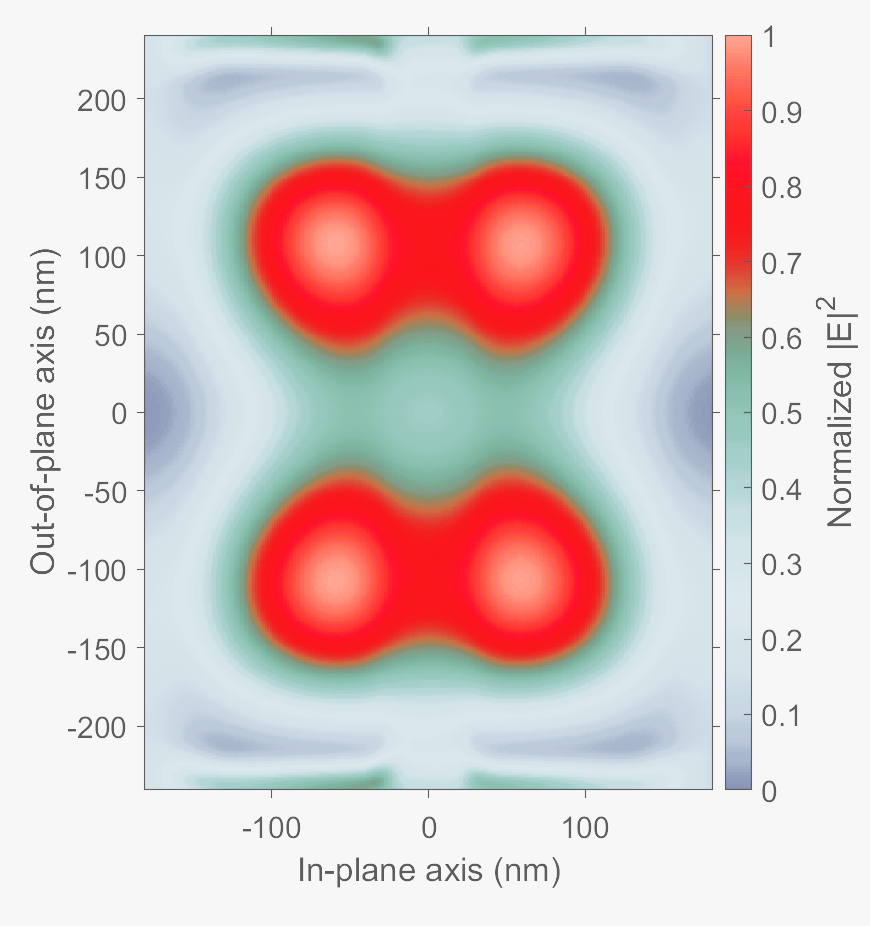}
  \vspace{2pt}\\
  \small\textbf{(b) TM, 90\textdegree}
\end{minipage}
\caption{Normalised electric field intensity $|E|^2$ at resonance,
$n_\mathrm{bg}=1.00$, cross-section through the mid-plane of the
bars (x--z plane).
\textbf{(a)} TE channel at 863.3\,nm: intensity peaks sharply in
the 60\,nm gap\rev{, as two narrow lobes running the height of the bars
against the two gap walls,} and within the bar volume\rev{, where two
broader secondary lobes appear}, indicating strong
field overlap with the analyte region and the \ce{MoO3} fill.
\textbf{(b)} TM channel at 960.1\,nm: intensity distributes in a
four-lobe pattern confined within the bars\rev{, above and below the
mid-height,} with reduced gap
occupation\rev{ and a local minimum at the centre of the gap}.
\rev{The weaker overlap of the TM mode with the gap region is
consistent with the smaller effect of the fill on that channel.}}
\label{fig:efields}
\end{figure}

Fano fits to the $\alpha$-\ce{MoO3} device at $n_\mathrm{bg}=1.00$
(Fig.\,\ref{fig:fano}) extract the resonance parameters.
The TE channel at $\lambda_0=863.3$\,nm fits with $Q=87$, asymmetry
parameter $q=-0.841$, FWHM\,$=9.89$\,nm, and $R^2=0.983$.
The TM channel at $\lambda_0=960.1$\,nm fits with $Q=31$,
$q=+0.384$, FWHM\,$=31.32$\,nm, and $R^2=1.000$.
The Fano lineshape arises from interference between the discrete
resonant mode and the broadband Fabry--P\'{e}rot
background.\cite{Limonov2017Fano,Miroshnichenko2010Fano}
The two resonances \rev{are} polarization-orthogonal
\rev{modes} of the bar pair\rev{, and
Fig.\,\ref{fig:efields} shows that they occupy the gap region to
different extents.
The two intensity maps differ in topology as well as in magnitude.
In the TE channel the maximum lies inside the 60\,nm gap, as two narrow
lobes running the height of the bars against the two gap walls, with
two broader secondary lobes within the bar volume.
In the TM channel the intensity forms four lobes in a two-by-two
arrangement within the bars, above and below the mid-height, with a
local minimum at the centre of the gap.
The TE mode therefore overlaps the gap fill and the analyte region far
more strongly than the TM mode does, which is consistent with the fill
having the larger effect on the TE channel.
Localised field enhancement of this kind is the quantity of interest in
related numerical studies of subwavelength
structures.\cite{Ramola2026JPHOT}}
The opposite signs of $q$ reflect different phase relationships
between the discrete and continuum paths under orthogonal
polarizations.
\rev{The ratio $Q_\mathrm{TE}/Q_\mathrm{TM}=2.85$ is reported as a
property of the simulated device. No scaling relation between the
axis-pair contrast and the two quality factors is proposed: calibrating
one against the TE channel would leave the TM channel as its only
independent test, which these data do not provide.}

\subsection{Dual-Channel Sensing Performance}

\begin{figure}[h]
\centering
\includegraphics[width=\textwidth]{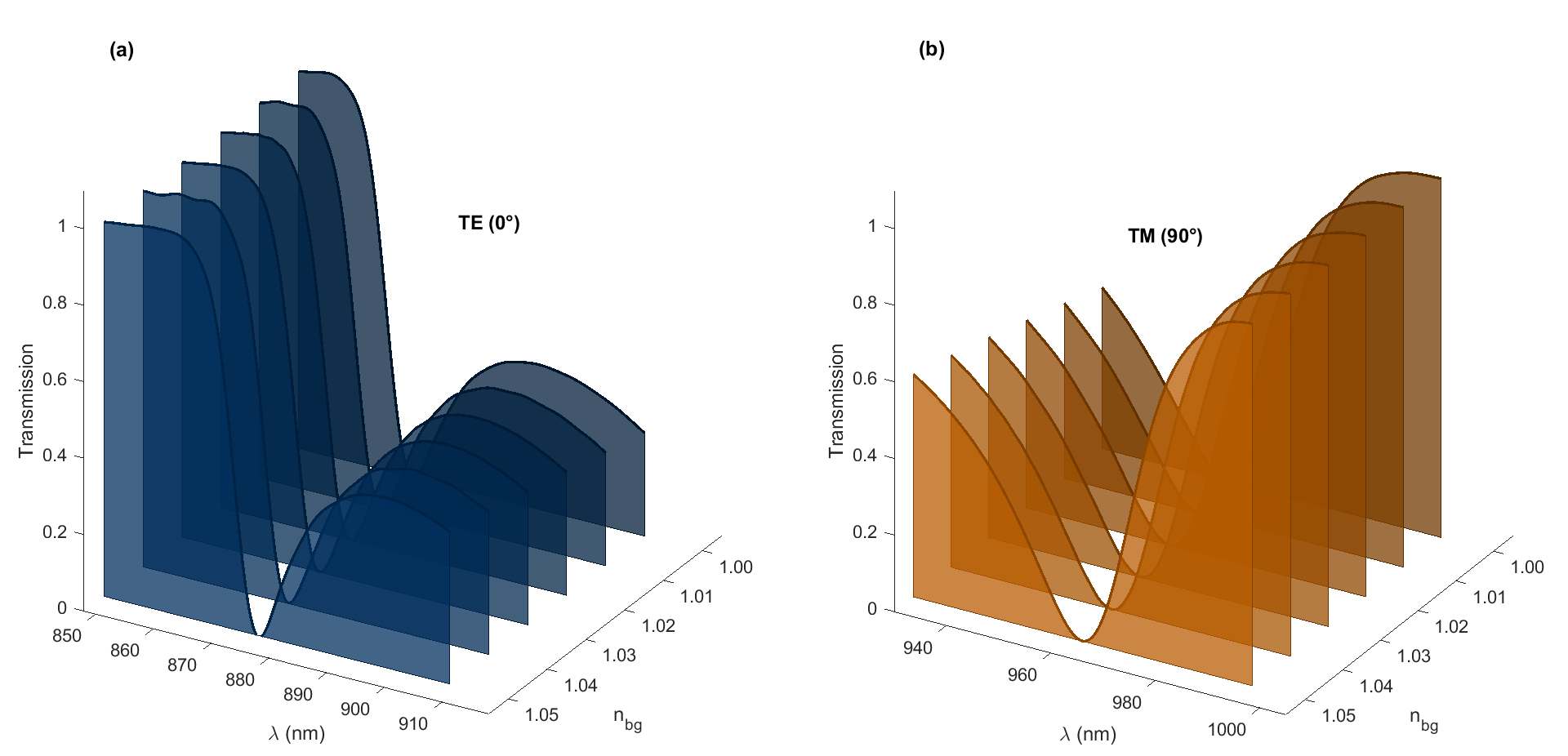}
\caption{$\alpha$-\ce{MoO3} transmission sweeps,
$n_\mathrm{bg}=1.00$--$1.05$.
\textbf{(a)} TE (860--915\,nm).
\textbf{(b)} TM (940--1010\,nm).
Both Fano resonances shift monotonically; the 97\,nm channel gap
prevents spectral overlap across the full RI range.}
\label{fig:waterfall}
\end{figure}

Fig.\,\ref{fig:waterfall} shows transmission sweeps across
$n_\mathrm{bg}=1.00$--$1.05$ for the $\alpha$-\ce{MoO3} device.
Both Fano resonances shift monotonically with no mode-crossing over
the 0.05\,RIU range.

\begin{figure}[h]
\centering
\includegraphics[width=0.6\textwidth]{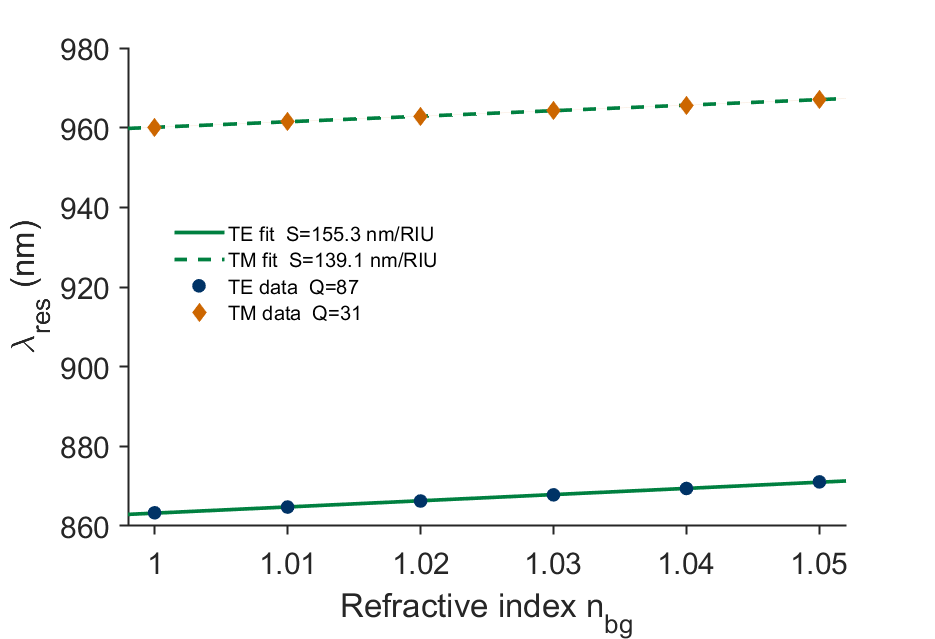}

\smallskip\noindent\textbf{(a)}\smallskip

\includegraphics[width=0.92\textwidth]{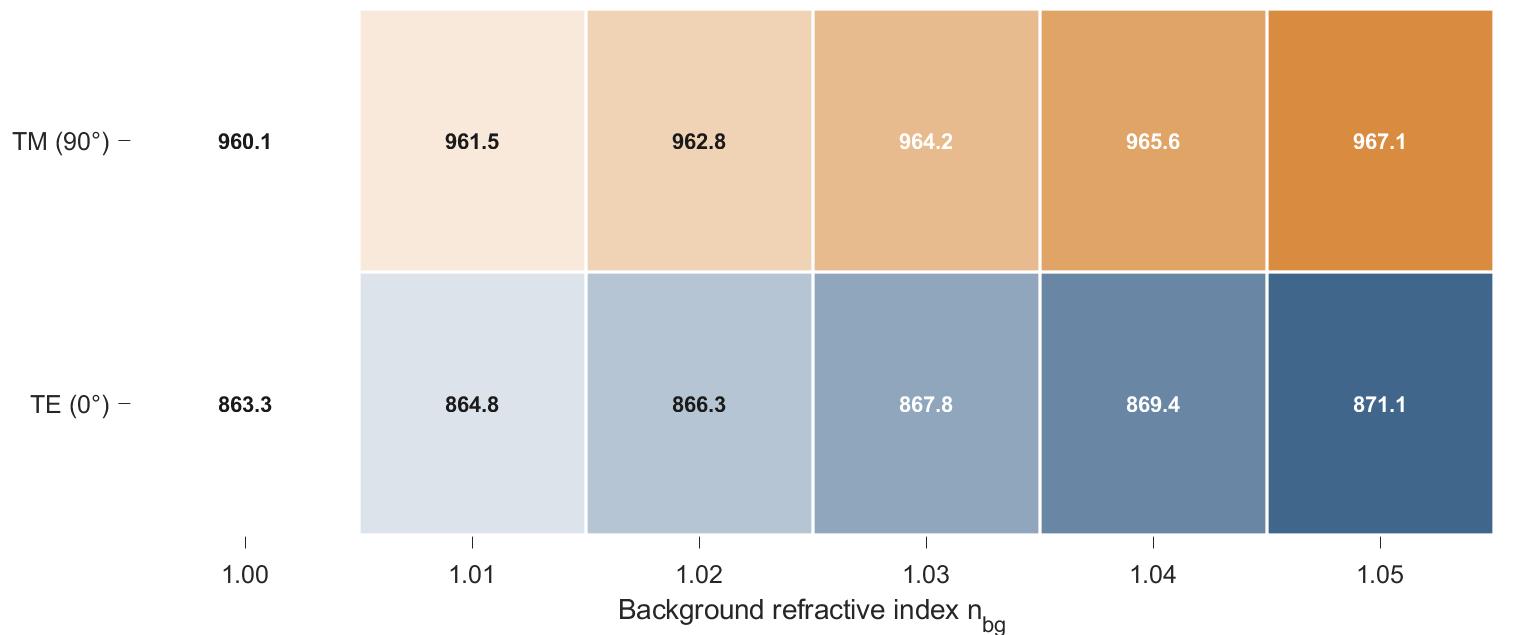}

\smallskip\noindent\textbf{(b)}
\caption{\textbf{(a)} Resonance wavelength vs.\ $n_\mathrm{bg}$:
$S_\mathrm{TE}=155.3$\,nm\,RIU$^{-1}$ and
$S_\mathrm{TM}=139.1$\,nm\,RIU$^{-1}$ ($R^2>0.999$).
\textbf{(b)} Resonance wavelength map (nm).
TE (blue, 863--871\,nm) and TM (orange, 960--967\,nm) occupy
non-overlapping windows across the full analyte RI range.}
\label{fig:sensitivity}
\end{figure}

Linear regression on $\lambda_\mathrm{res}$ vs.\ $n_\mathrm{bg}$
(Fig.\,\ref{fig:sensitivity}a, $R^2>0.999$) gives
$S_\mathrm{TE}=155.3$\,nm\,RIU$^{-1}$ and
$S_\mathrm{TM}=139.1$\,nm\,RIU$^{-1}$.
Figure of merit and refractive index resolution are defined
as\cite{White2008FOM,Soler2017LoD}
\begin{equation}
\mathrm{FOM}=\frac{S}{\mathrm{FWHM}},\qquad
\rev{\delta n}=\frac{\delta\lambda_{\min}}{S},
\label{eq:metrics}
\end{equation}
with $\delta\lambda_{\min}=0.01$\,nm (10\,pm spectrometer
resolution assumed).
Results: $\mathrm{FOM}_\mathrm{TE}=15.71$\,RIU$^{-1}$,
$\mathrm{FOM}_\mathrm{TM}=4.44$\,RIU$^{-1}$,
\rev{$\delta n_\mathrm{TE}$}$=6.44\times10^{-5}$\,RIU, and
\rev{$\delta n_\mathrm{TM}$}$=7.19\times10^{-5}$\,RIU.
\rev{These are ideal refractive index resolutions set by the assumed
instrument, not experimental limits of detection, and the previous
version used the latter term in error.
A more conservative estimate follows from requiring a shift of one
tenth of the resonance linewidth to be resolved, which gives
$6.4\times10^{-3}$\,RIU for TE and $2.3\times10^{-2}$\,RIU for TM;
reporting both makes clear that the sub-$10^{-4}$\,RIU figures are
instrument-limited idealisations.}
The TE channel outperforms TM in FOM because
\rev{its Fano linewidth is narrower at a similar sensitivity; the
small difference in ideal resolution between the two channels follows
from their sensitivities alone, since $\delta\lambda_{\min}$ is the
same for both}.
Both channels remain below $10^{-4}$\,RIU \rev{on the ideal measure}.
The resonance wavelength map (Fig.\,\ref{fig:sensitivity}b) shows
non-overlapping spectral windows across all tested analyte indices.

\begin{figure}[tbp]
\centering
\includegraphics[width=\textwidth]{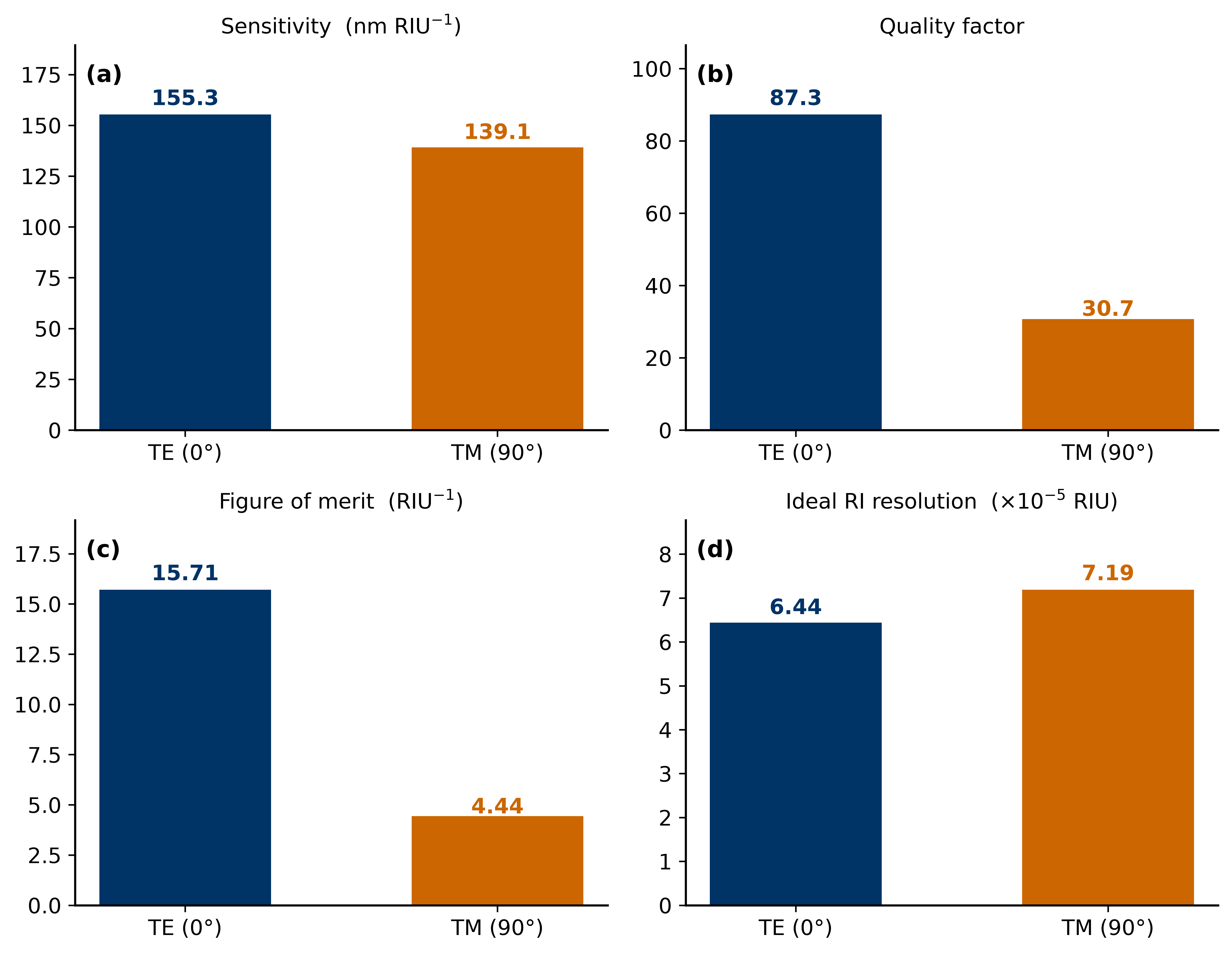}
\caption{Sensing metrics for TE (blue) and TM (orange).
\textbf{(a)} Sensitivity. \textbf{(b)} Quality factor.
\textbf{(c)} Figure of merit. \textbf{(d)} \rev{Ideal refractive
index resolution} ($\times10^{-5}$\,RIU; lower is better).}
\label{fig:metrics}
\end{figure}

Table\,\ref{tbl:comparison} contextualises these results.
\rev{The comparison identifies whether each entry reports measured or
simulated performance, since the values reported here are numerical.}
The TE FOM of 15.71\,RIU$^{-1}$ is competitive with single-channel
\ce{TiO2} and Si devices of comparable geometry.
The TM FOM of 4.44\,RIU$^{-1}$ \rev{follows from its broader
linewidth at a similar sensitivity}.

\begin{table}[tbp]
\centering
\caption{Selected all-dielectric metasurface RI sensors.
Y: Yes, N: No. Values from cited references.
\rev{Method: E, experimental; S, simulation; --, not determined here.}}
\label{tbl:comparison}
\small
\setlength{\tabcolsep}{5pt}
\begin{tabular}{llcccll}
\toprule
Material & Resonance & $\lambda$\,(nm)
  & $S$\,(nm\,RIU$^{-1}$) & FOM\,(RIU$^{-1}$)
  & \rev{Method} & Dual Channel \\
\midrule
Si           & q-BIC     & $\sim$900  & 241 & 84 & \rev{--} & N  \cite{Li2025quasiBIC}     \\
Si           & q-BIC     & $\sim$900  & 216 & 56 & \rev{S} & N  \cite{Jing2023quasiBIC}   \\
Si           & Fano      & $\sim$1064 & 147 & 30 & \rev{S} & N  \cite{Zhang2018OL}        \\
Si           & Toroidal  & $\sim$1220 & 180 & 22 & \rev{S} & N  \cite{Zhao2024TD}         \\
Si           & Dual Fano & $\sim$1550 & 145 & 18 & \rev{--} & Y \cite{Wang2024dualFano}   \\
TiO$_2$     & GMR       & $\sim$1064 & 102 & 8  & \rev{S} & N  \cite{Wu2022TiO2GMR}      \\
Si$_3$N$_4$ & q-BIC     & $\sim$780  & 98  & 12 & \rev{--} & Y \cite{Yang2025Si3N4}      \\
Si           & Perm-qBIC & $\sim$1060 & 190 & 40 & \rev{--} & N  \cite{Yang2025epsilonqBIC}\\
\midrule
\textbf{$\alpha$-MoO$_3$}
  & \textbf{Fano (TE)} & \textbf{863}
  & \textbf{155} & \textbf{15.71} & \rev{\textbf{S}}
  & \multirow{2}{*}{\textbf{Y [This work]}} \\
\textbf{$\alpha$-MoO$_3$}
  & \textbf{Fano (TM)} & \textbf{960}
  & \textbf{139} & \textbf{4.44} & \rev{\textbf{S}} \\
\bottomrule
\end{tabular}
\end{table}

\subsection{Dual-Polarization \rev{Response to Simultaneous Readout}}

\begin{figure}[tbp]
\centering
\includegraphics[width=0.62\textwidth]{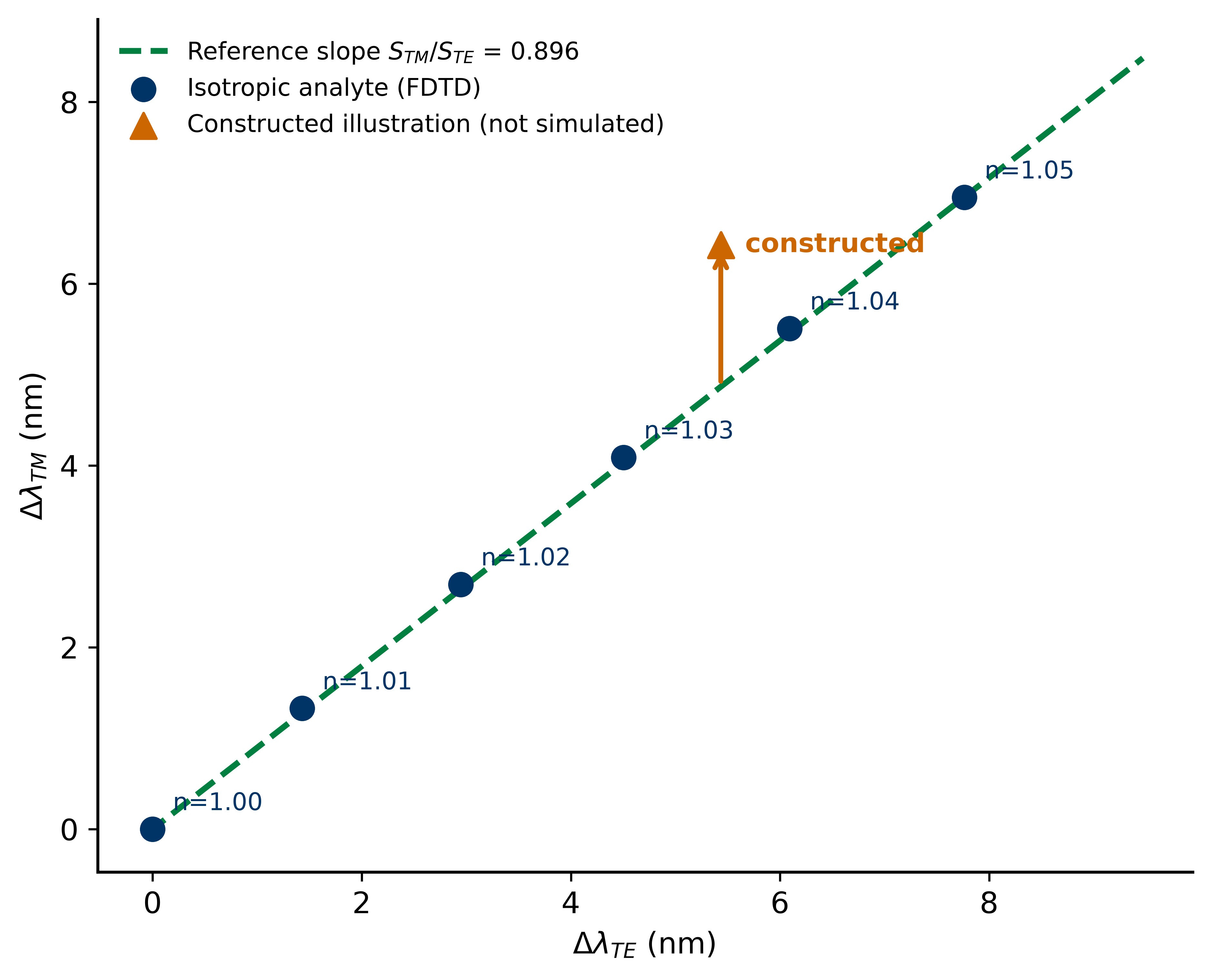}
\caption{\rev{Polarization response}: $\Delta\lambda_\mathrm{TM}$
vs.\ $\Delta\lambda_\mathrm{TE}$ for $n_\mathrm{bg}=1.00$--$1.05$.
Isotropic analytes (blue circles) fall on slope 0.896 (dashed).
\rev{The orange marker is a constructed illustration of an off-slope
point, not a simulated anisotropic analyte.}}
\label{fig:fingerprint}
\end{figure}

For an isotropic analyte, both channels respond to the same
$\Delta n$, fixing the shift ratio at
$\Delta\lambda_\mathrm{TM}/\Delta\lambda_\mathrm{TE}=0.896$
regardless of concentration or absolute RI.
\rev{This ratio is the ratio of the two channel sensitivities,
$S_\mathrm{TM}/S_\mathrm{TE}$, and the gap material changes it: the
same geometry filled with amorphous \ce{Sb2S3} gives 0.620.
Two fills do not establish how the ratio depends on the fill in
general, and we report the comparison rather than a relationship.}
Fig.\,\ref{fig:fingerprint} \rev{plots all six simulated analyte
indices in the $(\Delta\lambda_\mathrm{TE},\Delta\lambda_\mathrm{TM})$
plane.
Because the reference slope is obtained by regression on those same
points, their agreement with it is not an independent test; what the
six points support is the linearity of each channel, with
$R^2>0.999$.}
The isotropic slope is a sensor constant, not an analyte property.
An anisotropic analyte \rev{could produce unequal} TE and TM \rev{shifts},
displacing the
$(\Delta\lambda_\mathrm{TE},\Delta\lambda_\mathrm{TM})$ point from
the reference slope.
\rev{The point marked in Fig.\,\ref{fig:fingerprint} illustrating
this was constructed at a fixed displacement and is not the result of
a simulation; it is retained only as an illustration and is labelled
accordingly.
Whether the pair of measured shifts determines analyte birefringence
and optic-axis orientation uniquely is not addressed by these data,
since the response may also depend on analyte thickness, surface
coverage and inter-channel crosstalk, and we make no claim of
parameter retrieval here.}

\subsection{\rev{Fano asymmetry as a second observable}}

\rev{The asymmetry parameter $q$ of eqn~\eqref{eq:fano} measures the
relative weight of the direct and resonant scattering pathways and is
returned by every lineshape fit.
It carries a consistent signature in this device: $q$ is negative in the
TE channel and positive in TM for both gap fills, and its magnitude is
set by the fill, being larger for the isotropic material in both
channels ($-1.047$ and $+0.608$ for amorphous \ce{Sb2S3} against
$-0.841$ and $+0.384$ for $\alpha$-\ce{MoO3} at
$n_\mathrm{bg}=1.00$).}

\begin{figure}[tbp]
\centering
\includegraphics[width=\textwidth]{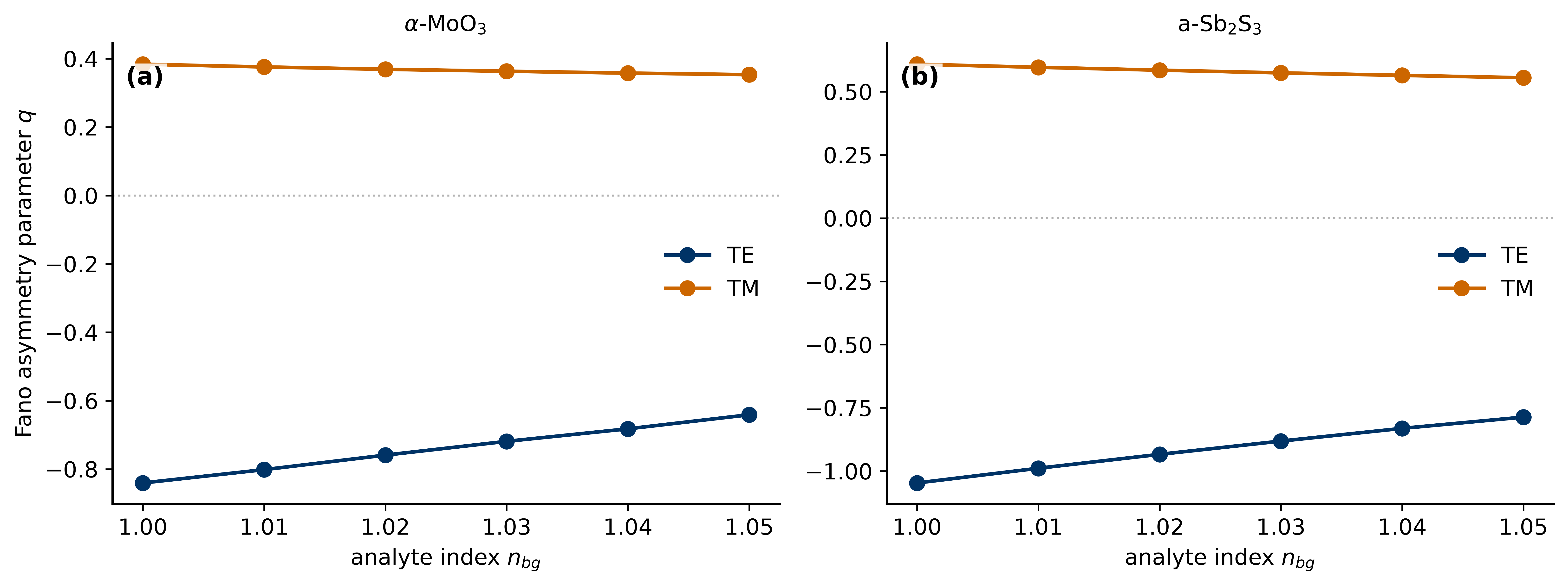}
\caption{\rev{Fano asymmetry parameter $q$ against analyte index.
\textbf{(a)} $\alpha$-\ce{MoO3} fill. \textbf{(b)} Amorphous \ce{Sb2S3}
fill. $q$ has opposite sign in the two channels and varies linearly with
$n_\mathrm{bg}$ in each ($R^2>0.99$).}}
\label{fig:fanoq}
\end{figure}

\rev{Across the analyte range studied, $q$ varies linearly with
$n_\mathrm{bg}$ and does so with opposite sign in the two channels
(Fig.\,\ref{fig:fanoq}).
For the $\alpha$-\ce{MoO3} device $\mathrm{d}q/\mathrm{d}n =
+3.99$\,RIU$^{-1}$ in TE and $-0.61$\,RIU$^{-1}$ in TM, with $R^2$ of
0.9997 and 0.9906 respectively.
We record this as an observation the present data support, with one
qualification: $q$ and $\lambda_0$ are obtained from the same
four-parameter fit and are therefore correlated, so $q$ is not an
independent measurement channel in the statistical sense.
Whether it can be used as one is not addressed by these data.}

\subsection{\rev{Lineshape response of the two channels}}

\begin{figure}[t]
\centering
\includegraphics[width=\textwidth]{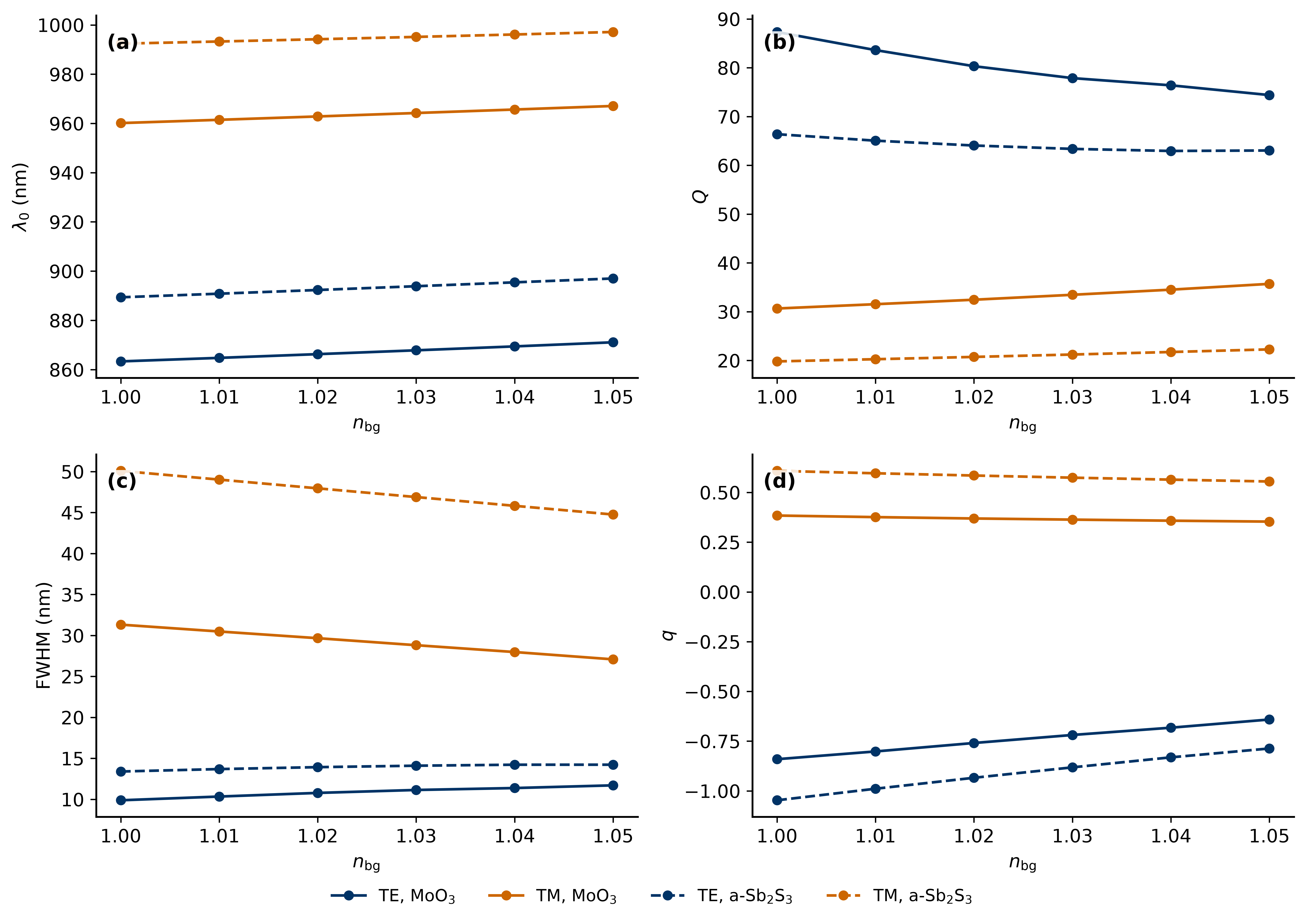}
\caption{\rev{Fano lineshape parameters against analyte index for both
gap fills. \textbf{(a)} Resonance wavelength. \textbf{(b)} Quality
factor. \textbf{(c)} Linewidth. \textbf{(d)} Asymmetry parameter.
Solid lines: $\alpha$-\ce{MoO3}; dashed: amorphous \ce{Sb2S3}. The two
channels respond in opposite senses in $Q$, in linewidth and in $q$,
while both resonance positions shift to longer wavelength.}}
\label{fig:lineshape}
\end{figure}

\rev{Tracking all four Fano parameters across the analyte range, rather
than the resonance position alone, shows a structure that holds for both
gap fills (Fig.\,\ref{fig:lineshape}).
The two resonance wavelengths both increase, giving the positive
sensitivities of Section~3.3.
The two quality factors move in opposite senses: $Q_\mathrm{TE}$ falls
from 87.3 to 74.4 across the range while $Q_\mathrm{TM}$ rises from 30.7
to 35.7, with the corresponding linewidths broadening and narrowing
respectively.
The same opposition is present with the amorphous \ce{Sb2S3} fill, where
$Q_\mathrm{TE}$ falls from 66.4 to 63.0 and $Q_\mathrm{TM}$ rises from
19.8 to 22.3.
The asymmetry parameter moves toward zero in both channels, approaching
it from opposite sides.}

\rev{The two channels are therefore complementary in every lineshape
parameter except the resonance position, and the ordering is unchanged
by the choice of gap material.
We report this as a characteristic of the device that the present
simulations support; the mechanism behind the opposed trends is not
established here.}

\subsection{\rev{Design rules}}

\rev{Comparing the two fills across all four channels
(Table~\ref{tbl:design}) supports two design rules, stated here as
observations from these simulations rather than as the outcome of a
formal optimisation, which was not performed.}

\begin{table}[tbp]
\centering
\caption{\rev{Simulated performance of all four channels.
$\delta n$ is the ideal refractive index resolution at an assumed
10\,pm spectrometer resolution.}}
\label{tbl:design}
\small
\begin{tabular}{llccccc}
\toprule
\rev{Gap fill} & \rev{Channel} & \rev{$S$ (nm\,RIU$^{-1}$)} & \rev{$Q$}
  & \rev{FWHM (nm)} & \rev{FOM (RIU$^{-1}$)} & \rev{$\delta n$ (RIU)} \\
\midrule
\rev{$\alpha$-\ce{MoO3}} & \rev{TE} & \rev{155.3} & \rev{87.3} & \rev{9.89}  & \rev{15.71} & \rev{$6.44\times10^{-5}$} \\
\rev{$\alpha$-\ce{MoO3}} & \rev{TM} & \rev{139.1} & \rev{30.7} & \rev{31.32} & \rev{4.44}  & \rev{$7.19\times10^{-5}$} \\
\rev{a-\ce{Sb2S3}}       & \rev{TE} & \rev{153.1} & \rev{66.4} & \rev{13.40} & \rev{11.43} & \rev{$6.53\times10^{-5}$} \\
\rev{a-\ce{Sb2S3}}       & \rev{TM} & \rev{94.9}  & \rev{19.8} & \rev{50.09} & \rev{1.89}  & \rev{$1.05\times10^{-4}$} \\
\bottomrule
\end{tabular}
\end{table}

\rev{The first rule follows from the TE rows.
Changing the gap fill from $\alpha$-\ce{MoO3}, whose in-plane
principal indices are 2.411 and 2.092, to amorphous \ce{Sb2S3} at
$n=2.90$ changes the TE sensitivity only from 155.3 to
153.1\,nm\,RIU$^{-1}$, about one percent, while the TE linewidth
changes by more than a third.
Sensitivity is governed by the overlap of the resonant mode with the
analyte region, and therefore by the geometry and the period, and is
comparatively insensitive to the choice of gap material; the fill acts
principally on the linewidth.
The spectral separation of the two channels likewise remained near
100\,nm for both fills.}

\rev{The second rule follows from the definitions in
eqn~\eqref{eq:metrics}.
Since $\mathrm{FOM}=S/\mathrm{FWHM}$, narrowing the resonance
linewidth at fixed sensitivity raises both the quality factor and the
figure of merit, so those two are not competing objectives.
The ideal refractive index resolution behaves differently: with
$\delta n=\delta\lambda_{\min}/S$ evaluated at a fixed instrumental
$\delta\lambda_{\min}$, it is set by the sensitivity alone and is
unaffected by the linewidth.
It is the linewidth-based resolution introduced in Section~3.3 that
narrowing the linewidth improves.
The trade-off lies instead in the reliability of peak tracking, as the
crystalline \ce{Sb2S3} case in Section~3.1 illustrates: raising the
fill index narrows the available linewidths but brings additional
resonances into the operating window, and the useful range ends where
the spectrum ceases to be single-peaked.
Geometry and gap material can therefore be treated as largely
separable design variables.}

\subsection{\rev{Application context}}

\rev{The device operates as a two-channel bulk refractive index
sensor with simulated sensitivities of 155.3 and
139.1\,nm\,RIU$^{-1}$ and ideal refractive index resolutions below
$10^{-4}$\,RIU in both channels.
The relevance of this operating range is illustrated by label-free
work in which normal and cancerous cell populations are distinguished
through refractive index differences within the 1.33--1.37
window.\cite{Shakya2026PCF}
The advantage of a two-channel readout in such a setting is
referencing: a bulk index change common to both polarizations
displaces the two resonances along the fixed ratio of Section~3.4,
whereas a process that alters the near-field distribution differently
for the two polarizations displaces the operating point away from
that ratio.
A single-resonance sensor cannot separate these two contributions
without an additional reference arm.}

\rev{A second setting is optically ordered biological films.
Lipid bilayers, collagen fibres and oriented nucleic acid monolayers
present different refractive indices along and across their ordering
axis; a single-polarization sensor records a weighted average of the
two, whereas simultaneous TE and TM readout retains a dependence on
that ordering.
A third is material characterisation: because the two channels sample
different in-plane directions of the material occupying the gap, the
structure could provide a fixed test platform for probing the in-plane
optical response of a small volume of a van der Waals crystal without
requiring a new resonator geometry for each material.
We have not carried out such a characterisation here.}

\rev{We state the limits of these remarks explicitly.
The present study is entirely numerical, and the analytes modelled
here are homogeneous index media rather than specific biomolecular
species.
Quantitative prediction of performance against a named biomarker
would require surface binding and coverage models outside the scope of
this work, and no biosensing result is reported here.}

\subsection{\rev{Modelling Assumptions and Tolerances}}

\rev{This section sets out the assumptions on which the simulations
rest.
It does not report fabrication, and the routes noted below are
possibilities drawn from the literature rather than processes
reported here.}

Both constituent materials have established deposition routes.
\ce{TiO2} nanobars at these dimensions \rev{would be} defined by
electron-beam lithography and reactive-ion etching of ALD or
sputtered films, which reach the $n=2.35$ index used
here.\cite{Profijt2011ALD,Garcia2021TiO2}
The 60\,nm gap has a height-to-width aspect ratio near four (250\,nm
tall, 60\,nm wide), within the range reached for \ce{TiO2} etching in
reported all-dielectric sensors.\cite{Abbas2020TiO2,Wu2022TiO2GMR}
Biaxial van der Waals crystals, including $\alpha$-\ce{MoO3}, are
integrated into nanophotonic structures by mechanical exfoliation or
chemical vapour deposition, with flake thicknesses from a few tens of
nanometres to bulk.\cite{Caldwell2019}
\rev{The simulations assume the gap is completely filled with a
single crystal of uniform orientation.
We do not claim that a flake transfer would achieve this: filling a
60\,nm wide, 250\,nm deep trench completely is not a straightforward
consequence of a transfer step, and conformal growth followed by
planarisation appears the more plausible of the two routes.
Complete single-crystal filling should be read as an idealisation
adopted for the numerical study.}

The resonance wavelengths are set by the bar dimensions and period, so
lithographic deviations in $W$, $g$, and $P$ shift $\lambda_\mathrm{TE}$
and $\lambda_\mathrm{TM}$.
The \rev{axis-pair contrasts of the fill are} fixed by the crystal and
independent of the patterned dimensions.
\rev{The relative effect of the fill on the two channels would
therefore be expected to persist under dimensional variation, with
the absolute resonance positions and sensitivities set per device by}
a one-time calibration against known-index media, as in any RI sensor.
\rev{We have not quantified this numerically, and the tolerance to
incomplete filling, reduced fill depth and dimensional variation
remains to be evaluated.}

Crystal-axis orientation relative to the bar pair sets which axis pair
each polarization samples.
The design aligns the $b$-axis with $x$ (across the gap) and the
$a$-axis with $y$ (along the bars).
A misalignment of the in-plane axes by an angle $\phi$ introduces an
off-diagonal permittivity component
$\varepsilon_{xy}=\rev{(\varepsilon_\beta-\varepsilon_\alpha)}\sin\phi\cos\phi$
that couples the TE and TM gap fields.
\rev{This component is non-zero for any $\phi$ other than
0\textdegree{} and 90\textdegree{} and removes both mirror planes of
the device, so that $\hat{x}$ and $\hat{y}$ cease to be
eigenpolarizations of the zeroth-order transmission.
For an isotropic fill $\varepsilon_\beta=\varepsilon_\alpha$ and it
vanishes at every angle, at any refractive index, so polarization
conversion under in-plane rotation is a signature of biaxiality that
an isotropic material cannot reproduce regardless of its index.
The simulations reported here were performed at $\phi=0$\textdegree{}
only, so the magnitude of this coupling is not quantified.}
Crystal-axis orientation can be read before transfer by
polarisation-resolved Raman mapping, which resolves the in-plane axes
of $\alpha$-\ce{MoO3} from the anisotropy of its phonon modes.

\section{Conclusions}

\rev{Gap-fill crystal anisotropy acts as a tuning parameter for the
two polarization channels of a} \ce{TiO2} nanobar-pair metasurface.
\rev{With an $\alpha$-\ce{MoO3} fill the device yields}
$Q_\mathrm{TE}=87$ and $Q_\mathrm{TM}=31$ with 97\,nm channel
separation and sub-$10^{-4}$\,RIU \rev{ideal refractive index
resolutions} in both channels, \rev{for the fixed single-unit-cell
geometry studied here}.

The \ce{Sb2S3} comparison is the key control: amorphous \ce{Sb2S3}
($n=2.90$) and crystalline \ce{Sb2S3} ($n=3.52$) bracket and exceed
the \ce{MoO3} principal values\rev{, and the amorphous phase yields a
quality-factor ratio of 3.35 against 2.85 for the biaxial fill.
Unequal TE and TM quality factors are therefore a property of the
in-plane anisotropy of the resonator geometry rather than of the
crystal symmetry of the fill, so an unequal pair of quality factors is
not by itself evidence of material anisotropy, and the device retains its mirror
planes when the principal axes are aligned with the structural axes,
so the resonances are not symmetry-broken quasi-BICs}.
Crystalline \ce{Sb2S3} additionally activates multi-mode TM
interference that degrades peak trackability, making it unsuitable
even as a high-index isotropic fill for this geometry.

\rev{What the biaxial fill provides at fixed geometry is control over
how the response is partitioned between the two channels.
Because each polarization samples a different principal index, one
material substitution moves the two resonance positions and the two
linewidths in different proportions and changes the ratio of the two
channel sensitivities, which is 0.896 for $\alpha$-\ce{MoO3} against
0.620 for amorphous \ce{Sb2S3}.
Sensitivity itself is governed by the geometry and is nearly
unchanged between fills, so the two design variables act on different
metrics and can be treated separately.}

The polarization \rev{response} (isotropic slope\,$=0.896$) extends
the device function beyond scalar RI sensing.
Biologically ordered films such as lipid bilayers, collagen fibres
\rev{and} oriented DNA \rev{would be expected to give a
polarization-dependent response that a single-polarization sensor
cannot resolve.
No such film was simulated here, and whether the resulting pair of
shifts determines analyte birefringence and axis orientation uniquely
is not established by this work}.

\rev{Two questions are left open by this work: the relationship
between the in-plane axis contrast and the two quality factors, which
requires a continuous sweep of the anisotropy at fixed fill-to-bar
contrast; and the magnitude of the polarization conversion under
in-plane rotation of the crystal axes, which vanishes identically for
any isotropic fill.
All results reported here are numerical, and experimental
verification remains to be carried out.}


\section*{\rev{Author contributions}}
\rev{\textbf{Shoumik Debnath:} conceptualisation, methodology,
software, formal analysis, investigation, visualisation, writing --
original draft, writing -- review and editing.
\textbf{Sudipta Saha:} supervision, validation, writing -- review and editing, project administration.}

\section*{Conflicts of interest}
There are no conflicts to declare.

\section*{Data availability}
The Ansys Lumerical FDTD simulation files used in this work are
available at
\url{https://github.com/debnath-shoumik/TiO2-MoO3-DualChannel-RI-Sensor};
\rev{raw spectra, fitting routines and figure-generation scripts will
be released at the same location on publication, and} additional data
are available from the corresponding author upon reasonable request.

\section*{Acknowledgements}
The authors acknowledge the computational resources of the
Department of Electrical and Electronic Engineering,
Bangladesh University of Engineering and Technology.
\rev{The authors thank the reviewers, whose comments identified an
error in the analysis of the \ce{Sb2S3} data and led to a revised
interpretation of the results.}

\bibliography{refs}
\end{document}